\documentclass[aps,prl,reprint,superscriptaddress]{revtex4-1}
\usepackage{amsmath,amssymb}
\usepackage{multirow}
\usepackage{graphicx}
\usepackage{epstopdf}
\usepackage{tabularx, booktabs}
\usepackage[titletoc,toc,title]{appendix}
\usepackage[colorinlistoftodos]{todonotes}
\usepackage{pbox}
\usepackage{siunitx}
\newcolumntype{Y}{>{\centering\arraybackslash}X}
\usepackage[caption=false,position=b,singlelinecheck=off,font=normalsize,labelfont=bf,justification=justified]{subfig}
\usepackage{hyperref}
\usepackage{adjustbox} 
\usepackage[version=4]{mhchem}
\hypersetup{
    colorlinks=true,
    linkcolor=blue,
    filecolor=black,      
    urlcolor=blue,                                                            
    citecolor=blue,}
\graphicspath{ {Figures/} }
\usepackage{xr}
\usepackage[normalem]{ulem}
\usepackage{xcolor}

\begin{document}
\title{The Propagating Nature of Single- and Multimagnons in \ce{LaCrO_3}}
\author{Masoud Lazemi}
\affiliation{Materials Chemistry and Catalysis, Debye Institute for Nanomaterials Science, Utrecht University, Universiteitsweg 99, 3584 CG Utrecht, The Netherlands.}
\affiliation{MESA+ Institute for Nanotechnology, University of Twente, P.O. Box 217, Enschede, 7500 AE, The Netherlands.}
\author{Fabian J. Mohammad}
\affiliation{Materials Chemistry and Catalysis, Debye Institute for Nanomaterials Science, Utrecht University, Universiteitsweg 99, 3584 CG Utrecht, The Netherlands.}
\author{Ellen M. Kiens}
\affiliation{MESA+ Institute for Nanotechnology, University of Twente, P.O. Box 217, Enschede, 7500 AE, The Netherlands.}
\author{Yorick A. Birkh\"{o}lzer}
\affiliation{MESA+ Institute for Nanotechnology, University of Twente, P.O. Box 217, Enschede, 7500 AE, The Netherlands.}
\author{Paras Mehta} 
\affiliation{Helmholtz Zentrum Berlin, Photon Science Division, 15 Albert-Einstein-Straße, 12489 Germany}
\author{Emma van der Minne} 
\affiliation{MESA+ Institute for Nanotechnology, University of Twente, P.O. Box 217, Enschede, 7500 AE, The Netherlands.}
\author{Stefano Agrestini} 
\affiliation{Diamond Light Source, Harwell Campus, Didcot, OX11 0DE, UK.}
\author{Frank M.F. de Groot}
\email[]{Frank M.F.deGroot@uu.nl}
\affiliation{Materials Chemistry and Catalysis, Debye Institute for Nanomaterials Science, Utrecht University, Universiteitsweg 99, 3584 CG Utrecht, The Netherlands.}
\author{Hebatalla Elnaggar}
\email[]{hebatalla.elnaggar@helmholtz.berlin.de}
\affiliation{Helmholtz Zentrum Berlin, Photon Science Division, 15 Albert-Einstein-Straße, 12489 Germany}
\affiliation{Institute of Mineralogy Physics of Materials and Cosmochemistry, CNRS, Sorbonne University, 4 Place Jussieu, Paris 75005, France.}

\date{\today}

\begin{abstract}
The dispersion of an excitation is commonly used to infer its microscopic character: dispersive modes are associated with propagation, whereas nondispersive features are often interpreted as localized. This distinction becomes less straightforward for composite excitations, whose spectral maxima can remain nearly dispersionless even when they are built from propagating quasiparticles. Here we use momentum- and temperature-dependent resonant inelastic X-ray scattering (RIXS) to reveal the origin of multimagnon excitations in antiferromagnetic LaCrO$_3$. Whereas the single magnon disperses strongly, the two- and three-magnon features remain nearly flat. On approaching the Néel temperature, thermal population of the magnon band produces an anti-Stokes branch and opens an additional three-magnon channel involving the creation of two magnons and annihilation of one. Its energy and momentum dependence are reproduced by a thermally weighted multimagnon joint density of states calculations. These results establish thermal population as a means of exposing the propagating constituents hidden within nondispersive composite excitations, and position RIXS as a probe of multimagnon dynamics and, ultimately, magnon–magnon interactions.
\end{abstract}
\maketitle

The dispersion relation of an elementary excitation provides one of the most direct links between spectroscopy and the dynamics of the excitation. A strongly dispersive mode is generally associated with a propagating excitation, whereas a nearly momentum-independent feature is often interpreted as localized. This distinction becomes considerably less straightforward for composite excitations. When several quasiparticles are created simultaneously, the momentum transferred by the probe can be distributed among the constituent excitations, such that their combined spectral weight may appear nearly nondispersive even when each constituent is itself propagating. Establishing the microscopic character of such apparently flat excitations therefore requires information beyond their observed dispersion.

Magnetic systems offer a particularly useful setting in which to address this problem. Their elementary collective excitations, magnons, propagate through the lattice with a dispersion determined by the underlying exchange interactions. The resonant inelastic X-ray scattering (RIXS) process can, however, create several magnons simultaneously, producing multimagnon excitations whose energy and momentum dependence reflect combinations of states throughout the Brillouin zone \cite{Nag2020,Elnaggar2023,Li2023}. These excitations contain information that is inaccessible from the single-magnon dispersion alone, including higher-order spin correlations, magnon--magnon interactions, decay processes, and possible correlated or bound states \cite{deGroot1998,Betto2021,Nag2022Quadrupolar,Kumar2022,Pal2023,Singh2025Bimagnon}. Distinguishing a continuum assembled from propagating magnons from a localized spin excitation or a genuine multimagnon bound state remains a central challenge in interpreting such spectra.

Inelastic neutron scattering (INS) has established much of our understanding of magnetic excitations by giving direct access to the momentum- and energy-dependent spin correlation function. Although multimagnon scattering can also occur in INS, its spectral weight is generally much weaker than that of the dominant single-magnon response. RIXS provides a complementary route to higher-order spin excitations \cite{deGroot2024RIXS}. At transition-metal absorption edges, the resonant intermediate state and the strong core-hole spin--orbit and exchange interactions allow a single scattering event to generate multiple spin excitations \cite{Ament2010,Betto2021,Elnaggar2023,Li2023}. RIXS can therefore access single-, double-, triple-, and higher-order magnon channels with appreciable spectral weight.

An additional dimension is attained by temperature. At low temperature, RIXS predominantly probes processes in which magnetic excitations are created from the ground state. Upon heating, however, thermally populated magnons become available in the initial state. RIXS can then annihilate as well as create magnons, giving rise to anti-Stokes scattering and to additional multimagnon channels that are absent at low temperature. Such processes give a direct probe of the constituent quasiparticles: an excitation involving the annihilation of a thermally occupied magnon necessarily retains information about the momentum and energy of the underlying propagating magnon states. Temperature-dependent RIXS can therefore reveal the microscopic origin of apparently nondispersive multimagnon features that cannot be obtained from their dispersion alone.

Here, we exploit this approach to determine the character of multimagnon excitations in the antiferromagnet LaCrO$_3$. Momentum-resolved Cr $L_3$-edge RIXS reveals a strongly dispersive single-magnon excitation together with nearly nondispersive two- and three-magnon features. Upon increasing the temperature towards the N\'eel temperature, the magnetic excitations soften but remain visible as paramagnons, while thermal population of the magnon band produces a dispersive anti-Stokes branch. Most importantly, an additional three-magnon scattering channel emerges in which two magnons are created while a thermally populated magnon is annihilated. We reproduce the energy and momentum dependence of this excitation using a thermally weighted multimagnon joint density of states calculation. The observation of this thermally activated channel demonstrates that the apparently nondispersive multimagnon features are assembled from propagating magnons and establishes temperature-dependent RIXS as a means of resolving the microscopic dynamics hidden within composite spin excitations.


We studied a thin-film of the perovskite antiferromagnetic insulator LaCrO$_3$ by tracking its magnetic excitation spectrum across the Néel transition using RIXS. Measurements acquired below and just above $T_{\mathrm N}$ allow us to distinguish the temperature evolution of the dispersive single-magnon response from that of the higher-order magnetic channels. By combining single-site full-multiplet calculations with linear spin-wave theory and multimagnon calculations, we establish a description of the observed excitations and examine the extent to which the multimagnon response can be understood in terms of thermally populated, propagating spin excitations.

LaCrO$_3$ is well suited to addressing this question. The half-filled $t_{2g}^{3}$ configuration of Cr$^{3+}$ largely quenches the orbital angular momentum, making LaCrO$_3$ a comparatively clean platform for investigating spin excitations. Moreover, it is a prototypical G-type antiferromagnet in which the Cr$^{3+}$ ions ($3d^3$, $S=3/2$) order below a Néel temperature, $T_{\mathrm N}$, reported to lie between approximately 275 and \SI{320}{\kelvin} \cite{Sahu2007}. This experimentally accessible transition temperature facilitates measurements both below and above the magnetically ordered phase. The antiferromagnetic order exhibits a small spin canting induced by the Dzyaloshinskii--Moriya interaction, resulting in a weak ferromagnetic moment below $T_{\mathrm N}$ \cite{Dzyaloshinsky1958WeakFerromagnetism,Zhou2010}. 

\begin{figure}[h!]
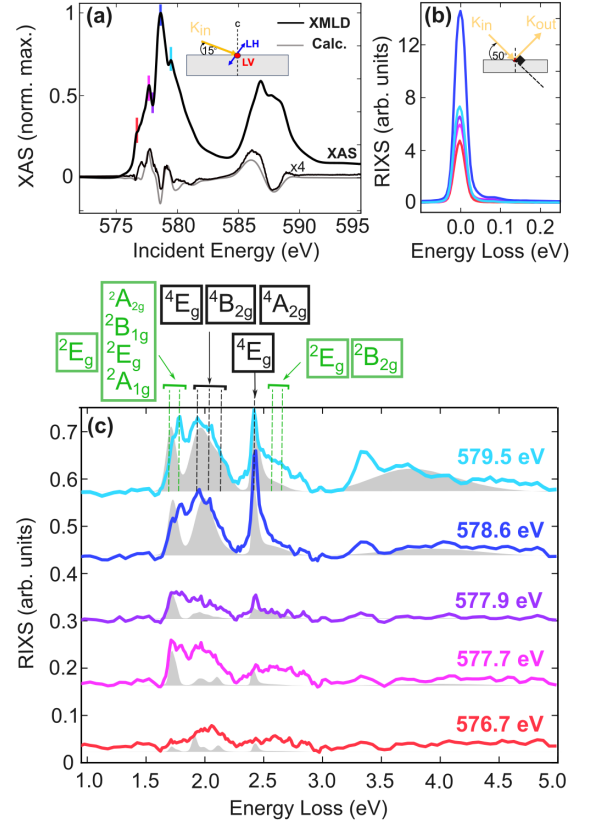

    \begin{center}
    \adjustimage{max size={1.3\linewidth}}{/RIXSMap_15K.png}
    \end{center}
	\caption{Cr \emph{2p} XAS and RIXS of LaCrO$_3$/Nb:SrTiO$_3$ thin film. (a) Experimental Cr  L$_{2,3}$-edge XAS measured with linear horizontal (LH) and linear vertical (LV) polarizations at \SI{15}{\kelvin}, along with the corresponding XMLD signal (scaled by a factor of 4 for clarity). The incident energies at which RIXS measurements were acquired at are indicated with colored lines. Theoretical XMLD spectra is shown in gray. (b) Cr \emph{2p3d} RIXS spectra acquired at an incident energies indicated by the arrow in panel (a) for scattering angle TTH = $\SI{90}{\degree}$ and using LV polarization. (c) Extended energy range of Cr \emph{2p3d} RIXS spectra. The calculated term symbols assigned to the RIXS spectra are shown where spin doublet states are color-coded in green and spin quadruplets in black.} \label{Fig:XAS_RIXSMap_15K}
\end{figure}

Because the magnetic anisotropy of LaCrO$_3$ is highly sensitive to epitaxial strain and film thickness, we determined the orientation of the Néel vector using angle-dependent X-ray magnetic linear dichroism (XMLD) at T = \SI{15}{\kelvin}. The measurements indicate that the Néel vector is oriented predominantly out of plane, consistent with previous results for LaCrO$_3$ films of comparable thickness \cite{Park2018,Sharma2021} (Full angular dependence of the XMLD is shown in the Supplemental Material Fig.~\ref{FigSI:CrXMLD_ThetaRotation}) and details about the growth of the film ans the characterization is provided in the Supplemental Material. The measured XMLD line shape also closely resembles that reported previously (Fig.~\ref{Fig:XAS_RIXSMap_15K}a) \cite{Park2018}. Complementary SQUID magnetometry, presented in the Supplemental Material Fig.~\ref{FigSI:SQUID}, agrees with the out of plane as the easy magnetization axis.

To support the interpretation of the XMLD spectrum, we performed ligand-field multiplet calculations, shown by the gray curve in Fig.~\ref{Fig:XAS_RIXSMap_15K}a using the code Quanty \cite{Haverkort2012Multiplet}. Further information about the calculation is provided in the Supplemental Material. The calculated line shape reproduces the principal features of the experimental spectrum reasonably and the values used are close to these reported in previous work \cite{Hunault2018Direct}. To facilitate comparison, the experimental XMLD amplitude was multiplied by a factor of four. This reduced experimental amplitude may arise, at least in part, from averaging over magnetic domains and the presence of oxygen vacancies which we infer from the reciprocal space mapping as shown in the Supplemental Material.
A notable feature of the present data is that the XMLD amplitudes and the fluorescence yield signal at the Cr $L_3$ and $L_2$ edges are comparable, unlike in many transition-metal systems in which the $L_3$ dichroism is substantially stronger. The pronounced $L_2$-edge dichroism motivates future polarization-dependent RIXS measurements at this edge, including RIXS-XMLD and RIXS-XMCD, where reduced saturation effects may offer an experimental advantage.

We next turn to the RIXS results shown in Figs.~\ref{Fig:XAS_RIXSMap_15K}b and c. To investigate the electronic excitation spectrum in greater detail, we acquired an incident-energy-dependent RIXS map with an intermediate energy resolution of $\Delta E=\SI{31.2}{\milli\electronvolt}$. In the low-energy-loss region below \SI{250}{\milli\electronvolt}, the spectra exhibit a shoulder associated with magnetic excitations, although the individual magnon contributions cannot be clearly resolved at this energy resolution. The first $dd$ excitations appear at approximately \SI{1.7}{\electronvolt} and extend to \SI{3}{\electronvolt}, followed by broader spectral features between 3.5 and \SI{5}{\electronvolt}. Within the 1.7--\SI{3}{\electronvolt} range, four groups of excitations can be distinguished.

These features are reproduced by ligand-field multiplet calculations for a tetragonally distorted CrO$_6$ cluster; further details of the calculations are provided in the Supplemental Material. In $D_{4h}$ symmetry, the Cr$^{3+}$ ground state is $^4B_{1g}$, derived from the $^4A_{2g}$ ground term in $O_h$ symmetry. The first two relatively sharp groups of $dd$ excitations are assigned predominantly to spin-doublet states, including the $^2E_g$ state and the tetragonally split $^2A_{2g}$, $^2B_{1g}$, $^2E_g$, and $^2A_{1g}$ multiplets. The subsequent group is dominated by transitions to the spin-quartet $^4E_g$, $^4B_{2g}$, and $^4A_{2g}$ states. These excitations are somewhat broader than the lower-energy doublet excitations, possibly reflecting additional splitting or broadening associated with magnetic exchange interactions. Most notably, the spectrum contains an intense excitation at approximately \SI{2.42}{\electronvolt}, which the calculation assigns predominantly to a transition to the $^4E_g$, accompanied by weaker contributions from the $^2E_g$ and $^2B_{2g}$ multiplets.

\begin{figure}[h!]
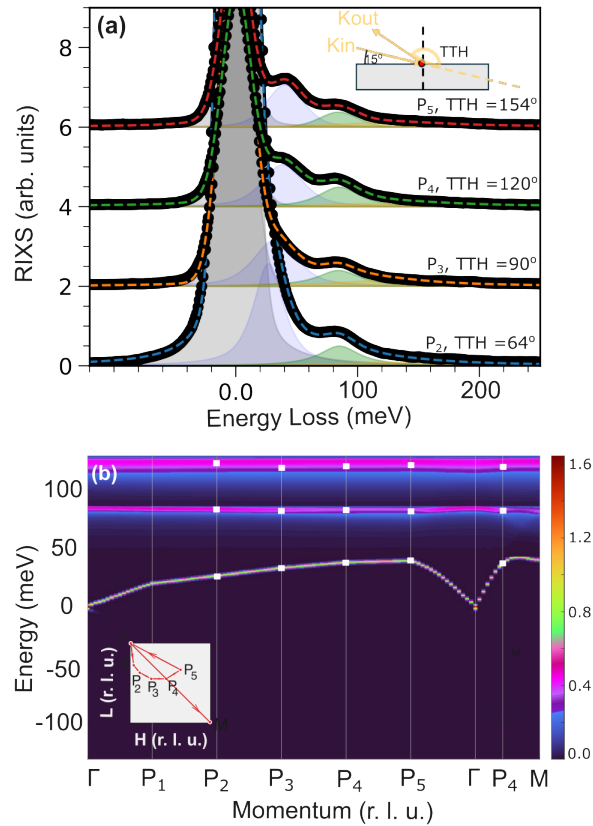

    \begin{center}
    \adjustimage{max size={1.3\linewidth}}{/Dispersion_15K.png}
    \end{center}
	\caption{(a) Magnon dispersion at the Cr L$_3$-edge ((E$_{\mathrm{in}}$ = \SI{578.6}{\electronvolt}) measured at \SI{15}{\kelvin}, with an incident angle ($\theta$) of 15$^\circ$ and scattering angles (TTH) of 154$^\circ$, 120$^\circ$, 90$^\circ$, and 64$^\circ$. Black circles represent the experimental data, and dashed lines show fits using pseudo-Voigt functions. Details of the fitting procedure are in the Supplementary Materials. (b) Linear spin-wave calculations of the single- and multimagnon excitations within the noninteracting approximation. The experimental data (square symbols) are overlaid on the calculated spectra, and the measurement path in the Brillouin zone is shown schematically in the inset. The points P$_1$–P$_5$ are not high-symmetry points; their coordinates are in the Supplementary Materials.} \label{Fig:Dispersion_15K}
\end{figure}

To resolve the low-energy magnetic excitations, we performed high-energy-resolution RIXS measurements ($\Delta E=\SI{20.4}{\milli\electronvolt}$) at $T=\SI{15}{\kelvin}$. The incident photon energy was fixed at \SI{578.6}{\electronvolt}, corresponding to the maximum of the XAS, and the incident angle was set to $\theta=15^\circ$. The momentum dependence of the excitations was mapped by varying the scattering angle, TTH, as shown in Fig.~\ref{Fig:Dispersion_15K}a.

The improved energy resolution reveals several distinct low-energy features. In the representative spectrum measured at TTH = 154$^\circ$, we identify a single-magnon excitation near \SI{40}{\milli\electronvolt} (blue), a two-magnon feature near \SI{84.5}{\milli\electronvolt} (green), and a three-magnon feature near \SI{124.5}{\milli\electronvolt} (orange). As shown in Fig.~\ref{Fig:Dispersion_15K}b, the energy of the single-magnon excitation varies clearly with momentum transfer. The measured dispersion is reproduced well by linear spin-wave calculations performed using the SpinW package \cite{Toth2015SpinW} and agrees well with inelastic neutron scattering data \cite{McQueeney2008}.

In contrast to the single-magnon excitation, the maxima of the two- and three-magnon features show no discernible dispersion within our experimental resolution. This behavior is also captured by our noninteracting multimagnon calculations \cite{Haverkort2010}, which produce nearly momentum-independent intensity maxima for both channels. Importantly, this does not necessarily imply that the constituent magnons are localized. For a multimagnon excitation, the measured momentum transfer corresponds only to the total momentum, which can be distributed among the individual magnons in many different ways. At each total momentum, the spectrum can sample a large number of magnon combinations throughout the Brillouin zone. The nearly flat maxima arise because many of these combinations accumulate near similar summed energies, producing pronounced peaks in the multimagnon joint density of states even though the individual magnons remain dispersive.

\begin{figure}[h!]
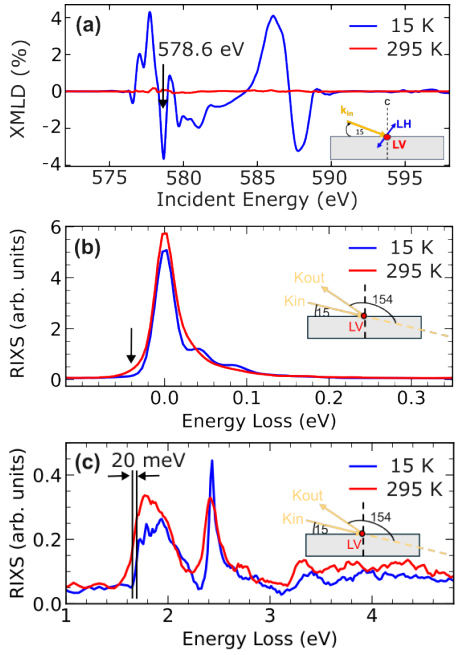

    \begin{center}
    \adjustimage{max size={1\linewidth}}{/XMLD_RIXS.png}
    \end{center}
	\caption{Cr \emph{2p} XAS and RIXS of LaCrO$_3$/Nb:SrTiO$_3$ thin film. (a) Experimental Cr \emph{2p} XMLD measured at \SI{15}{\kelvin} (blue) and at \SI{295}{\kelvin} (red). The measurement geometry is shown in the inset. (b) Cr L$_3$ edge (E$_{in}$ = \SI{578.6}{\electronvolt}) measured at \SI{15}{\kelvin} (blue) and at \SI{295}{\kelvin} (red) and zoomed in to see the magnon excitations. The incident angle (th) was 15$^\circ$ and at the scattering angles (TTH) of 154$^\circ$. (c) Extended energy range of panel (b) showing the $dd$ excitations.} \label{Fig:TempComparison}
\end{figure}

To investigate the evolution of the single- and multimagnon excitations near the Néel temperature, we repeated the measurements at $T=\SI{295}{\kelvin}$. Figure~\ref{Fig:TempComparison}a compares the Cr $L$-edge XMLD spectra measured at \SI{15}{\kelvin} (blue) and \SI{295}{\kelvin} (red). At \SI{295}{\kelvin}, the XMLD amplitude is strongly suppressed and approaches zero, consistent with the loss of long-range antiferromagnetic order near $T_{\mathrm N}$.

Figure~\ref{Fig:TempComparison}b compares the corresponding low-energy RIXS spectra. Upon heating, both the single- and multimagnon features shift to lower energies, revealing a clear thermal softening of the magnetic excitations. In addition, spectral weight develops on the energy-gain side, as indicated by the arrow. This anti-Stokes contribution arises when the RIXS process annihilates a thermally populated magnetic excitation and transfers its energy to the scattered photon. The increased energy-gain intensity gives a direct evidence for the thermal population of low-energy magnon excitations. We emphasize that the instrumental energy resolution was nearly identical at both temperatures, as demonstrated by fits to the elastic lines presented in the Supplemental Material Fig.~\ref{FigSI:ElasticFit}; the observed changes therefore cannot be attributed to a difference in experimental resolution. In addition, the dispersion of the anti-Stokes magnon peak is evident in Fig.~\ref{FigSI:RIXSDispersion}, where the high- and low-temperature RIXS spectra measured at different TTH values are overlaid. The high-temperature spectra cannot be reproduced by simply broadening the corresponding low-temperature spectra. If the observed changes arose solely from temperature-induced broadening, there would be no reason for the line shape on the energy-gain side of the elastic peak to evolve differently between TTH (=154$^\circ$) and (90$^\circ$), as observed in Fig.~\ref{FigSI:RIXSDispersion}. The momentum-dependent evolution of this additional spectral weight supports its assignment to a dispersive anti-Stokes magnon excitation.

Interestingly, the $dd$ excitations also exhibit a red-shift of approximately \SI{20}{\milli\electronvolt} at \SI{295}{\kelvin} compared to at \SI{15}{\kelvin}, as shown in Fig.~\ref{Fig:TempComparison}c. We attribute this shift primarily to a change in the screening where we observe that the values of the Slater integrals (F$^2_{dd}$ and F$^4_{dd}$) has to be increased to reproduce the shift as discussed in the Supplementary Materials. This shift can be seen throughout the full incident energy map (Fig.~\ref{FigSI:RIXSEnergyMaps}) and momentum transfer (Fig.~\ref{FigSI:RIXSDispersion}). Furthermore, the sharp peak at \SI{2.42}{\electronvolt} is clearly broadened which supports our assignment of the peak to the $^4$E$_g$ state. As this state comprises of eight degenerate microstates, coupling to phonons can lower the local symmetry and lift the degeneracy, resulting in a pronounced broadening of the excitation with increasing temperature.

\begin{figure*}[t]
    \centering
    \adjustimage{width=0.75\textwidth}{multimagnon_processes_V4.png}
    \caption{(a) Magnon dispersion at the Cr $L_3$ edge
    ($E_{\mathrm{in}} = \SI{578.6}{\electronvolt}$), measured at \SI{295}{\kelvin} with an incident angle $\theta = 15^\circ$ and scattering angles (TTH) of
    154$^\circ$, 120$^\circ$, and 90$^\circ$. Black circles represent
    the experimental data, and shaded peaks show pseudo-Voigt fits.
    Details of the fitting procedure are provided in the Supplementary
    Materials. (b) Linear spin-wave calculations of the single- and
    multimagnon excitations within the noninteracting approximation. The
    experimental data (square symbols) are overlaid on the calculated
    spectra. (c) Schematic spectrum to illustrates the corresponding spectral regions: the conventional $2\mathrm{M}$ and $3\mathrm{M}$ creation continua, the thermally activated $3\mathrm{M}$ ($++-$) feature, and near-zero-energy magnetic weight from the $2\mathrm{M}$ ($+-$) channel. The anti-Stokes single-magnon response directly evidences the thermal population required for the mixed creation and annihilation channels. The broad spectral envelopes represent maxima of multimagnon continua assembled from dispersive single magnons.}
    \label{Fig:Dispersion_295K}
\end{figure*}

Figure~\ref{Fig:Dispersion_295K}a shows fits to the high-temperature RIXS spectra. Despite the strong suppression of the XMLD signal and the associated loss of long-range antiferromagnetic order, magnetic excitations remain visible at \SI{295}{\kelvin}. Their persistence indicates substantial short-range spin correlations near $T_{\mathrm N}$; we therefore refer to these excitations as paramagnons.

At TTH = 154$^\circ$, the single-magnon energy decreases from approximately \SI{40.5}{\milli\electronvolt} at \SI{15}{\kelvin} to approximately \SI{27}{\milli\electronvolt} at \SI{295}{\kelvin}. The two- and three-magnon features undergo a corresponding softening, appearing near 52 and \SI{81}{\milli\electronvolt}, respectively. We also resolve an energy-gain, or anti-Stokes, feature at approximately $\SI{-27}{\milli\electronvolt}$ (purple in Fig.~\ref{Fig:Dispersion_295K}a), with the same absolute energy as the Stokes single-magnon excitation. This feature results from the annihilation of a thermally populated magnon during the RIXS process. Fits that omit this anti-Stokes contribution, presented in the Supplemental Material in Fig.~\ref{Fig:CrDispersionFits_T295K_compareFits}, fail systematically to reproduce the energy-gain side of the elastic peak. Moreover, this asymmetry cannot be explained by an asymmetric instrumental response: the elastic-line profile measured from carbon tape is symmetric and closely resembles the low-temperature elastic profile (see Fig.~\ref{FigSI:ElasticFit}).

Upon decreasing the momentum transfer, the Stokes and anti-Stokes single-magnon branches both disperse toward lower absolute energies, reaching approximately $\SI{\pm 22}{\milli\electronvolt}$ at TTH = 90$^\circ$, the smallest scattering angle investigated. At this momentum, however, an additional contribution near \SI{28}{\milli\electronvolt} is required to reproduce the RIXS spectrum (yellow in Fig.~\ref{Fig:Dispersion_295K}a). If this component is omitted, the fit instead produces an abrupt and nonphysical increase in the linewidth of the dispersive single-magnon feature, as demonstrated in Fig.~\ref{FigSI:CrDispersionFits_T295K_6Vs7} of the Supplemental Material.

To identify the origin of this additional contribution, we compare the measurements with finite-temperature spin-wave and multimagnon calculations. To account phenomenologically for the thermal softening of the magnon energy near \(T_{\mathrm N}\), the high-temperature dispersion was calculated using an effective spin value of \(S_{\mathrm{eff}}=1\). This reproduces the observed softening of the magnetic excitation energies reasonably well. Using this finite-temperature single-magnon spectrum, we calculate the multimagnon response following the same phase-space-convolution procedure employed at low temperature. The result is shown in Fig.~\ref{Fig:Dispersion_295K}b, together with the fitted experimental energies indicated by white squares.

The calculation reproduces both the dispersive single-magnon branch and the characteristic energies of the multimagnon features. It also produces the anti-Stokes single-magnon branch, in good agreement with the experiment. Most importantly, the calculation reveals an additional, nearly momentum-independent feature near \SI{28}{\milli\electronvolt}. At TTH = 154$^\circ$ and $120^\circ$, this feature overlaps with the dispersive single-magnon response and cannot be resolved separately. At TTH = 90$^\circ$, the single-magnon branch disperses to approximately \SI{22}{\milli\electronvolt}, allowing the nearly flat \SI{28}{\milli\electronvolt} contribution to become distinguishable, in excellent agreement with the fitted spectra.

The calculation identifies this feature as a thermally activated three-magnon channel (see Fig.~\ref{Fig:Dispersion_295K}c). At low temperature, the dominant three-magnon process creates three magnons, with an energy transfer of the form: $\omega=\omega_1+\omega_2+\omega_3$. At \SI{295}{\kelvin}, the finite thermal magnon population enables an additional process in which two magnons are created and one pre-existing magnon is annihilated:
$\omega=\omega_1+\omega_2-\omega_3$.
This $++-$ channel generates the nearly flat spectral maximum around \SI{28}{\milli\electronvolt}. The simultaneous observation of the anti-Stokes single-magnon branch confirms the thermal population required for this process. The agreement among the measured energy, momentum dependence, and thermal activation of the feature therefore provides strong evidence that the multimagnon response contains a continuum assembled from propagating, thermally populated spin excitations, rather than being exclusively a local multiple-spin-flip excitation.

An alternative interpretation of the nearly momentum-independent feature near \SI{28}{\milli\electronvolt} is a folded one-magnon branch arising from the reduced crystallographic symmetry of LaCrO$_3$. Such a branch is possible in principle because of the enlargement of the structural or magnetic unit cell folds magnons from different pseudocubic momenta onto the same reduced wave-vector. Several observations, however, make this interpretation unlikely. First, our mode-resolved spin-wave calculation reveals no one-magnon mode near \SI{28}{\milli\electronvolt} with appreciable dynamical spin-correlation weight at the relevant experimental momenta and in agreement with previous work \cite{McQueeney2008}, whereas the thermally activated three-magnon $++-$ calculation produces a nearly flat spectral ridge at this energy. Second, at low temperature we measured the momentum $P_1=(0.0181626,0,-0.1377369)$, where the acoustic magnon occurs near \SI{18}{\milli\electronvolt} and a sufficiently intense optical mode near \SI{28}{\milli\electronvolt} should therefore produce additional spectral weight or a detectable modification of the line shape. No such contribution is required to describe the low-temperature spectrum (see Fig. \ref{FigSI:CrDispersionFits_T15K}). Finally, if the \SI{28}{\milli\electronvolt} feature was a one-magnon branch arising from the reduced crystallographic symmetry, thermal balance would imply a corresponding anti-Stokes contribution near \SI{-28}{\milli\electronvolt} with an expected thermal intensity ratio of approximately $\exp[-28/(k_{\mathrm B}T)]\simeq0.33$ at \SI{295}{\kelvin}. No additional energy-gain feature attributable to such an optical mode is observed. The combined absence of a weighted one-magnon mode in the calculation, of the corresponding low-temperature spectral contribution, and of an additional anti-Stokes partner strongly disfavors this scenario. The observed energy, weak momentum dependence, and thermal activation are instead consistently reproduced by the three-magnon $++-$ channel, in which two magnons are created and one thermally populated magnon is annihilated.

An analogous process occurs in the two-magnon channel: one magnon can be created while another thermally populated magnon is annihilated, $\omega=\omega_1-\omega_2$.
Because the two energies can be similar, this $+-$ channel produces magnetic spectral weight close to zero energy transfer. Within the experimental resolution, such low-energy magnetic scattering overlaps with the elastic line and may contribute to the enhanced central intensity observed at \SI{295}{\kelvin}, together with thermally enhanced phonon and other quasielastic contributions.

In conclusion, we have used temperature- and momentum-dependent Cr $L$-edge RIXS to determine the microscopic character of single- and multimagnon excitations in the G-type antiferromagnet LaCrO$_3$. At \SI{15}{\kelvin}, we resolve a dispersive single-magnon branch together with nearly momentum-independent two- and three-magnon features. The single-magnon dispersion is well described by linear spin-wave theory, while multimagnon calculations show that the nearly flat spectral maxima emerges from the joint density of states of dispersive magnons. The absence of an apparent dispersion in a composite excitation therefore does not necessarily imply localization of its microscopic constituents.

On approaching $T_{\mathrm N}$, the magnetic excitations soften but remain observable despite the suppression of long-range antiferromagnetic order, revealing the persistence of short-range spin correlations. Thermal population of the magnon band is directly manifested by the emergence of a dispersive anti-Stokes single-magnon branch and, more importantly, opens additional multimagnon scattering pathways. We identified a three-magnon channel in which two magnons are created while one thermally populated magnon is annihilated. Its energy, momentum dependence, and thermal evolution are reproduced by finite-temperature multimagnon calculations, providing direct evidence that the apparently nondispersive multimagnon response is assembled from propagating spin excitations.

More generally, our results show that thermal occupation can be turned from a source of spectral broadening into a spectroscopic handle on the dynamics of composite excitations. Creation--annihilation channels provide access to the propagating quasiparticles hidden within apparently nondispersive spectral features and offer information that cannot be obtained from their dispersion alone. More broadly, exploiting thermally populated initial states may establish finite-temperature spectroscopy as a means of accessing quasiparticle interactions and higher-order correlations that remain inaccessible from conventional dispersion measurements.

\section{Author contributions}
\indent H.E. and F.M.F.d.G. conceived and designed the experiments. RIXS measurements were performed by M.L., H.E. and F.M.F.d.G. with experimental support from S.A.. H.E. analyzed the RIXS and XAS data with assistance of M.L. and P.M.. The thin film sample was prepared by M.L. with assistance from E.v.M. SQUID magnetometry measurements were carried out by H.E. with assistance from Y.K., and the resulting data were interpreted by H.E. Multiplet calculations were performed by H.E. and F.M.. H.E. performed the linear spin-wave calculations and the multimagnon calculations. Preliminary XRD measurements were conducted by M.L. and E.v.M., while complementary HRXRD and RSM were performed by E.M.K. HRXRD simulations were carried out by E.M.K. and Y.A.B., and the data were interpreted by M.L. Sample growth was supervised by C.B.. The overall project supervision was provided by H.E.. The manuscript was written by H.E. with input from all co-authors.

\section{Acknowledgment}
\indent We acknowledge Diamond Light Source for beam time on I21-RIXS beamline (Proposal ID: MM33336). We are thankful to Mizuki Furo, Naoki Ito, Atsushi Hariki, Ru-Pan Wang, Benjamin Lenz,  Mirian Garcia-Fernandez and Dmitri A. Tenne for the fruitful discussions. We thank Yannick Klein and David Hrabovsky for their assistance with the SQUID magnetometry measurements.  H.E. acknowledges the funding of the Helmholtz Association and HZB of her Investigator Group. M.L. and F.M.F.d.G received funding from the European Union’s Horizon 2020 research and innovation program under the Marie Skłodowska-Curie grant agreement No. 860553. M.L. thanks Shu Ni for fruitful discussions on sample preparation. M.L. is grateful to Dominic Post and Daniel Monteiro Cunha for their support in the lab at the University of Twente. 

\bibliographystyle{unsrt}
\bibliography{Reference}

\newpage
\onecolumngrid
\section{Methods}
\subsection{Pulsed laser deposition growth}

The epitaxial LaCrO$_3$ thin-film was grown by pulsed laser deposition (PLD) with insitu reflection high-energy electron diffraction (RHEED) monitoring at the MESA+ Institute for Nanotechnology, University of Twente, Netherlands. A Nb:SrTiO$_3$ (100) substrate, purchased from CrysTec GmbH and TiO$_2$-terminated was used for the deposition after preparing the substrate as indicated in Reference \cite{Koster1998}. The growth was carried out at an oxygen pressure of 0.04 mbar and a substrate temperature of 650~$^\circ$C, using a KrF laser with a fluence of 1.8 J cm$^{-2}$, a target-to-substrate distance of 5~cm, and at a repetition rate of 2 Hz. 

\subsection{Reflection high-energy electron diffraction}
\begin{figure}[h!]
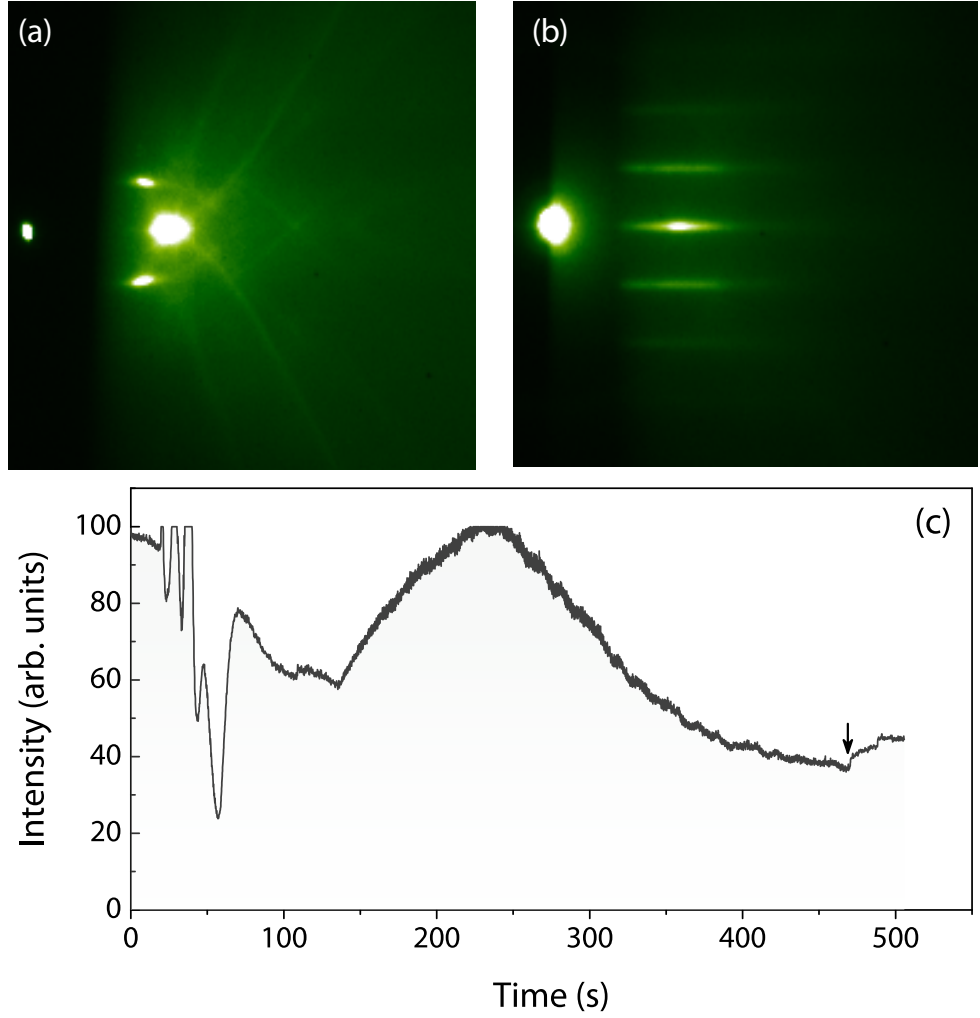

    \begin{center}
    \adjustimage{max size = 0.75\linewidth}{/FigS1pdf.pdf}
    \end{center}
	\caption{Insitu RHEED measurements confirm two-dimensional layer-by-layer growth and high crystalline quality of the LaCrO$_3$ thin film. (a) RHEED pattern of the Nb:SrTiO$_3$ (001) substrate prior to LaCrO$_3$ deposition, showing sharp streaks and Kikuchi lines indicative of an atomically flat and well-ordered surface. (b) Post-growth RHEED pattern of the LaCrO$_3$ thin film, displaying intense streaks in agreement with a smooth and crystalline surface. (c) Insitu RHEED specular intensity oscillations during LaCrO$_3$ deposition, exhibiting clear periodicity in agreement with two-dimensional layer-by-layer growth. The arrow marks the point at which growth is completed.}
	\label{Fig:RHEED}
\end{figure}

The growth was monitored insitu using reflection high-energy electron diffraction (RHEED). \emph{Insitu} RHEED measurements were performed during the growth of LaCrO$_3$ thin films to monitor surface crystallinity and growth mode. A 30-keV electron beam was directed at a grazing incidence angle to the substrate surface along the [100] azimuth of the SrTiO$_3$ (001) substrate.  Prior to deposition, the RHEED pattern of the substrate (Fig.~\ref{Fig:RHEED}a) exhibited sharp spot and Kikuchi lines, confirming an atomically flat, well-ordered surface. Upon initiation of growth, the specular RHEED intensity displayed pronounced oscillations as a function of time (Fig. \ref{Fig:RHEED}c), characteristic of a two-dimensional, layer-by-layer growth mode. The presence of several clear oscillations prior to amplitude damping indicates high surface mobility and smooth film nucleation. The arrow in the insitu RHEED intensity plot (Fig. \ref{Fig:RHEED}c) marks the point at which film growth was completed. The RHEED pattern acquired after growth retained intense, streaky features (Fig.~\ref{Fig:RHEED}b), indicating that the film surface remained smooth, and that the epitaxial growth mode was substantially preserved throughout the deposition process.

\subsection{X-ray reflectivity}

\begin{figure}[h!]
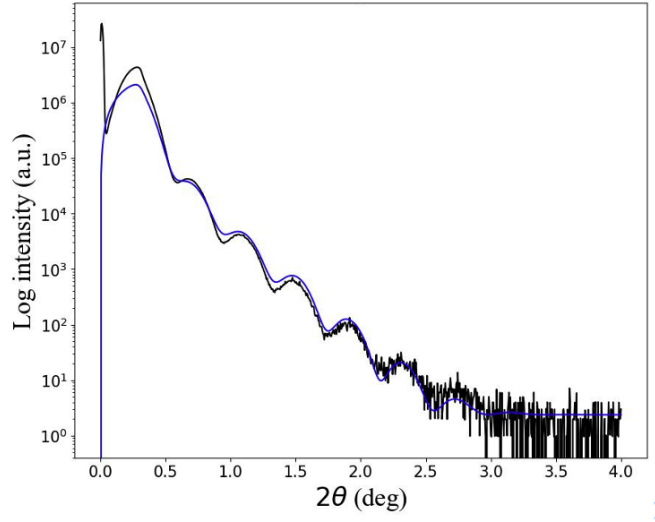

    \begin{center}
    \adjustimage{max size = 0.5\linewidth}{/XRR.png}
    \end{center}
	\caption{X-ray reflectivity manifests the smooth surface morphology and sharp interfaces of the LaCrO$_3$ thin film. X-ray reflectivity (XRR) profile of the LaCrO$_3$ thin film. The experimental data (black) and corresponding fit (blue) exhibit pronounced Kiessig fringes, confirming a smooth film surface and sharp interfaces.}
	\label{Fig:XRR}
\end{figure}

X-ray reflectivity (XRR) measurements yield Kiessig fringes spanning a wide angular range, confirming the presence of a smooth surface and a sharp film-substrate interface as shown in Fig.~\ref{Fig:XRR}. The fitted reflectivity profile shows excellent agreement with the measured data, corroborating the film thickness obtained from HRXRD ($\approx$10.4 nm). The sustained fringe visibility indicates low surface and interface roughness. The extracted film thickness ($\approx$10.4 nm) agrees with the HRXRD analysis.

\subsection{High-resolution X-ray diffraction}

\begin{figure}[h!]
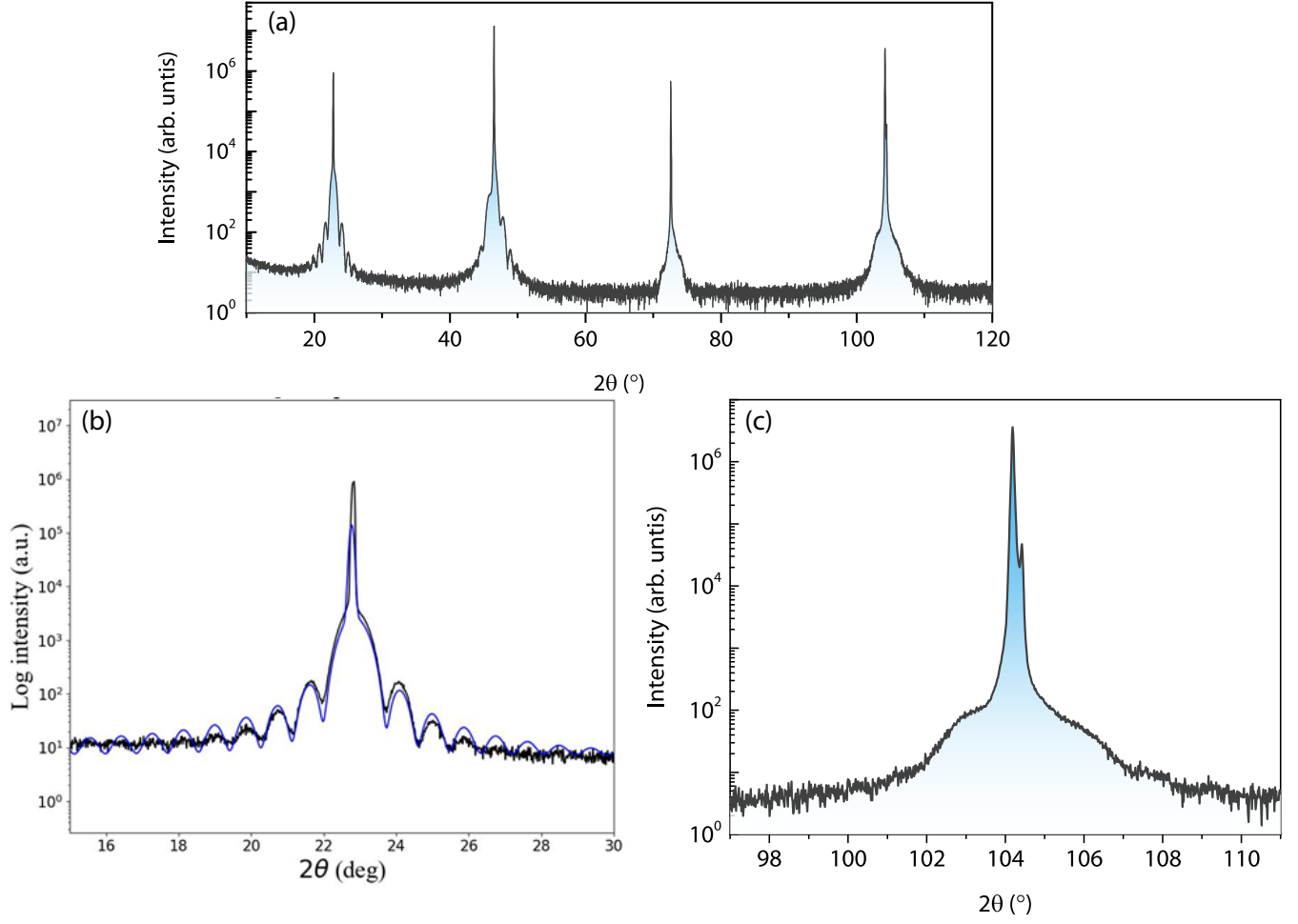

    \begin{center}
    \adjustimage{max size = \linewidth}{/HRXD.pdf}
    \end{center}
	\caption{High-resolution X-ray diffraction unravels the epitaxial growth, phase purity, and high crystalline quality of LaCrO3 thin films. (a) $\theta - 2\theta$ HRXRD scan of the LaCrO$_3$/Nb:SrTiO$_3$ (001), showing well-defined film and substrate reflections, indicative of phase purity and crystallographic alignment. (b) High-resolution scan of the LaCrO$_3$ (001) reflection with Laue fringe fitting (blue), yielding a film thickness of 10.4 nm. The presence of oscillations demonstrates high interface quality and smooth surface morphology. (c) HRXRD scan around the (004) reflection, showing that the film peak is indistinguishable from the sharp and intense substrate peak.}
	\label{Fig:HRXRD}
\end{figure}

$\theta - 2\theta$ high-resolution X-ray diffraction (HRXRD) scans (Fig.~\ref{Fig:HRXRD}a) show well-defined film peaks along with the substrate reflections, confirming phase purity and crystallographic alignment. A zoomed-in scan of the LaCrO3 (001) (Fig.~\ref{Fig:HRXRD}b) reveals pronounced Laue oscillations, from which a film thickness of approximately 10.4 nm was extracted. The presence of well-defined fringes further manifests the uniform film thickness and smooth interfaces. The high-order (004) reflection exhibit sharp and symmetric profiles which we interpret as the substrate peak (Fig.~\ref{Fig:HRXRD}c). Note that the double-peak feature is related to substrate imperfections and not related to the thin film. Even at this high-order reflection, the film and substrate peaks overlap due to their similar lattice parameters. Nevertheless, the observation of Laue fringes confirms the presence of a uniform film with good crystalline quality on the substrate.

\subsection{Reciprocal space mapping}

\begin{figure}[h!]
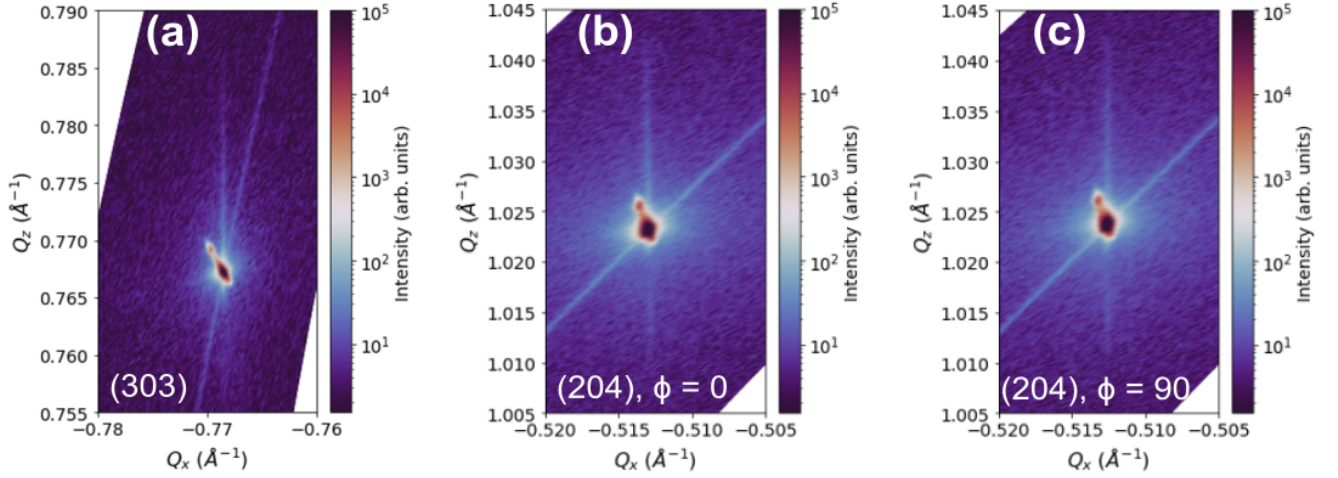

    \begin{center}
    \adjustimage{max size = \linewidth}{/RSM.png}
    \end{center}
	\caption{Reciprocal space mapping corroborates coherent epitaxial growth and lattice matching between LaCrO$_3$ and the SrTiO$_3$ substrate. (a) Reciprocal space map (RSM) of the LaCrO$_3$ thin film around the asymmetric (303) reflection. The substrate peak is the intense feature in the center of the RSM. The film signal corresponds to the vertical line near Q$_x$=-0.77, representing the film peak and related Laue fringes. The maximum intensity across this line is very close to and therefore indistinguishable from the substrate peak, as also confirmed by the symmetric XRD scans. (b) Reciprocal space map (RSM) of the LaCrO$_3$ thin film around the asymmetric (204) reflection acquired at $\phi= 0^o$ and (c) at $\phi= 90^o$. The alignment of the film and substrate peaks along both Q$_x$ and Q$_z$ demonstrates full coherent epitaxy and pseudomorphic growth, with in-plane and out-of-plane lattice parameters closely matching those of the substrate (a $\approx$ b $\approx$ c $\approx$ 3.905A).}
	\label{Fig:RSM}
\end{figure}

Reciprocal space mapping (RSM) centered around the (303) reflection (Fig.~\ref{Fig:RSM}a) reveals that the LaCrO$_3$ film peak is fully aligned with that of the SrTiO$_3$ substrate along both the in-plane (Q$_x$) and out-of-plane (Q$_z$) directions. We do not see a clear film peak above the substrate peak and only observe weak crystal truncation rod. Note that The side peak is an artifact from the X-ray mirror upstream from the monochromator of the Bruker D8 setup.
This indicates coherent epitaxy and substantiates that the in-plane lattice constants are pseudo-morphically matched to the substrate the out-of-plane lattice parameter also remains close to \SI{3.905}{\angstrom}. 

To further investigate structural coherence and domain formation, RSMs were acquired at the asymmetric (204) reflection $\phi= 0^o$ and $\phi= 90^o$ (Fig.~\ref{Fig:RSM}b and c). In both cases, a sharp, symmetric, and free of any observable splitting peak is seen along Q$_x$ and Q$_z$ and no clear substrate peak is observed suggesting that our film has a pseudo-cubic unit-cell. 

\begin{figure}[h!]
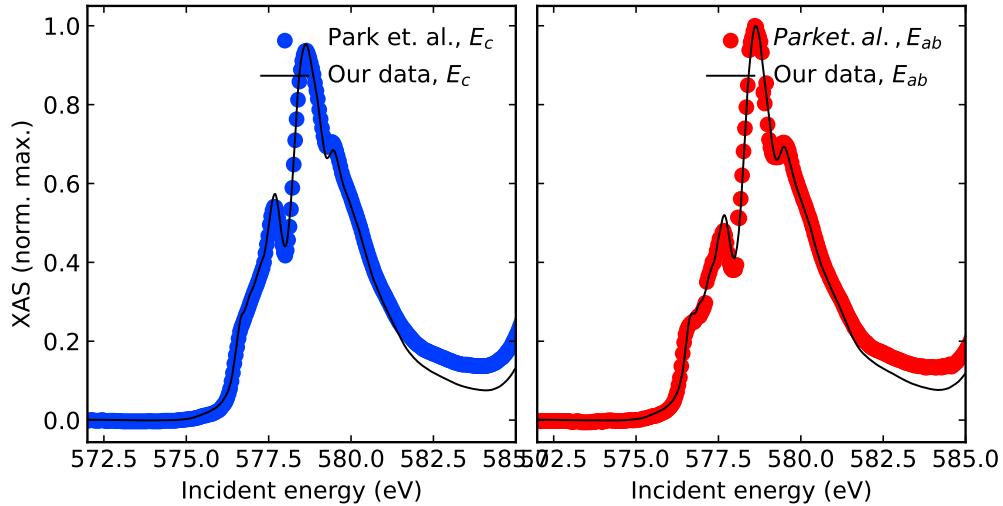

    \begin{center}
    \adjustimage{max size = 0.75\linewidth}{/CompareXAS.pdf}
    \end{center}
	\caption{Comparison of the Cr L$_3$ XAS spectra of our LaCrO$_3$ film with those reported in Ref. \cite{Park2018}. (a) XAS measured with the incident polarization parallel to the $c$ direction and (b) parallel to the $ab$ plane.}
	\label{Fig:XASCompare}
\end{figure}

The pseudo-cubic lattice parameter of bulk LaCrO$_3$ is expected to be \SI{3.885}{\angstrom}, suggesting that the film experiences in-plane tensile strain when grown on SrTiO$_3$. However, rather than the expected contraction of the out-of-plane lattice parameter to compensate for the in-plane tensile strain through the Poisson effect, we observe tensile strain along the out-of-plane direction as well. One possible explanation for this anomalous lattice expansion is the presence of oxygen vacancies, which are among the most common defects associated with shifts of XRD peaks toward lower $2\theta$ angles. The formation of charged oxygen vacancies can lead to an expansion of the lattice, thereby contributing to the observed out-of-plane lattice expansion. However, we have not observed any signature of oxygen deficiency when we compared the Cr XAS signal shown in Fig.~\ref{Fig:XASCompare} with work from reference \cite{Park2018} which suggests that such vacancies are not significant to change the average oxidation state of Cr in the film.

\clearpage

\section{Data analysis}
\subsection{Elastic line fitting}

\begin{figure}[h!]
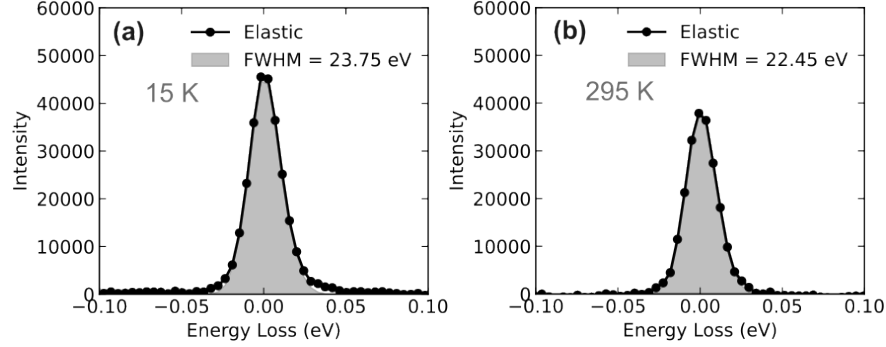

    \begin{center}
    \adjustimage{max size = 0.65\linewidth}{/ElasticLineFit.png}
    \end{center}
	\caption{Fitting of the elastic line measured on carbon tape measured at E$_{in}$ = 578.61 eV with LV polarization. (a) T = 15 K and. (b) T = 295 K.}
	\label{FigSI:ElasticFit}
\end{figure}

The elastic scattering was measured from a carbon-tape to determine the experimental resolution and the profile coming from the instrument. We used a single Gaussian function for the fit. Figure~\ref{FigSI:ElasticFit} shows these fits at T = 15 K and T = 295 K. The profile can be well captured using a symmetric Gaussian function and the full width half maximum (FWHM) of the elastic line was found to be 23.7~meV for the measurements at 15 K and 22.5~meV for the measurements at 295 K.

\subsection{RIXS fitting of the magnons dispersion at T = 15 K}

The low energy RIXS data were fitted using a Gaussian function to simulate the elastic scattering and three pseudo-voigt functions for the single-, double-, and triple-magnons. The FWHM of the single-, double-, and triple-magnon were determined from the RIXS data measured at scattering angle TTH = 154$^\circ$ as it showed the largest single-magnon energy allowing for a better quantification. 

\begin{figure}[h!]
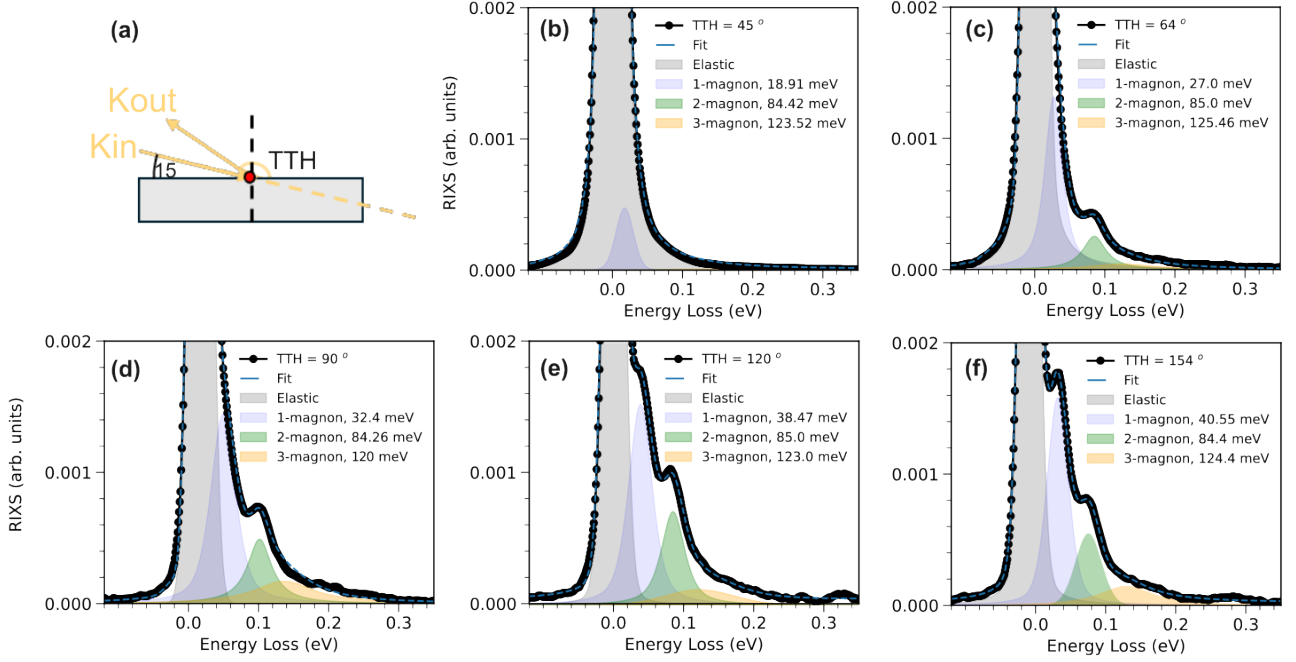

    \begin{center}
    \adjustimage{max size = \linewidth}{/CrDispersionFits_T15K.png}
    \end{center}
	\caption{Fitting of the Cr L$_3$-edge RIXS data measured at E$_{in}$ = 578.61 eV with incident X-ray angle of 15$^\circ$ using LV polarization at T = 15 K. (a) Sketch of the scattering geometry. (b-f) RIXS measurements with scattering angle TTH = 45$^\circ$, 64$^\circ$, 90$^\circ$, 120$^\circ$, and 154$^\circ$ correspondingly.}
	\label{FigSI:CrDispersionFits_T15K}
\end{figure}

The Cr L$_3$-edge RIXS data and the magnon fits measured at E$_{in}$ = 578.61 eV with incident X-ray angle of 15$^\circ$ using LV polarization at T = 15 K are shown in Fig.~\ref{FigSI:CrDispersionFits_T15K}. The fits are robust except for TTH = 45$^\circ$ as the elastic scattering becomes very strong approaching forward scattering. For this reason we have not used the RIXS results at TTH = 45$^\circ$. 

The relation between the scattering geometry and the RIXS momentum transfer, expressed in the pseudo-cubic unit cell, is summarized in Tab.~\ref{TabSI:TTH_HKL}. The measured point P$_4$ passes very close to the momentum path along the high-symmetry direction $\Gamma \rightarrow M$.
\begin{table}[h!]
    \centering
    \begin{tabular}{c|c|c|c}
        th ($^\circ$) & TTH ($^\circ$)& Name & q$_{pc}$ (r.l.u.)   \\
        \hline
        \hline
         15 & 45 & P$_1$ & (0.0181626, 0.0000000, -0.1377369)\\
         15 &  64 & P$_2$ & (0.0562724, 0.0000000, -0.1839678) \\
         15 &  90 & P$_3$ &  (0.1283731, 0.0000000, -0.2223044) \\
         15 & 120 & P$_4$ & (0.2223266, 0.0000000, -0.2223044)\\
         15 & 154 & P$_5$ &  (0.3123328, 0.0000000, -0.1660623)\\
        \hline
    \end{tabular}
    \caption{The RIXS momentum-transfer expressed in the pseudo-cubic unit cell of LaCrO$_3$.}
    \label{TabSI:TTH_HKL}
\end{table}

\subsection{RIXS fitting of the magnons dispersion at T = 295 K}

The low-energy RIXS spectra were fitted using a Gaussian function to model the elastic scattering and four pseudo-Voigt functions to account for the single-, double-, and triple-magnon excitations and the single anti-Stokes magnon contribution. In addition, an extra pseudo-Voigt function was included to describe a feature observed at $\approx 154$~meV in all high-temperature spectra, which we attribute to a phonon.

\begin{figure}[h!]
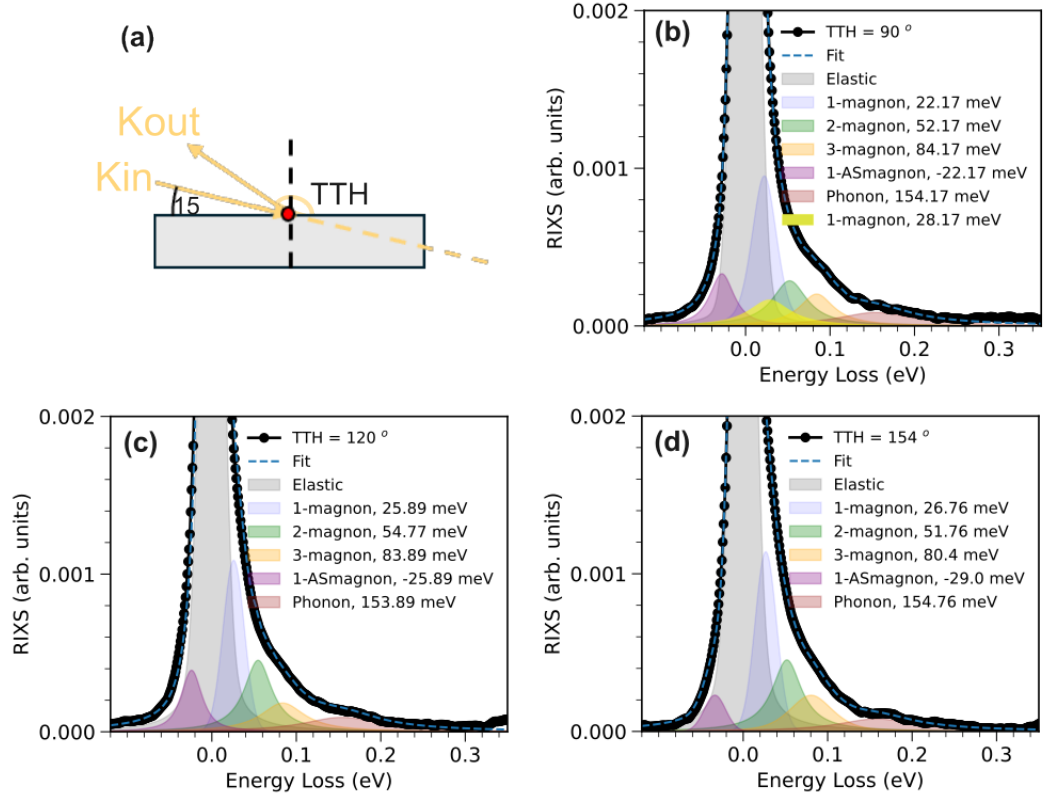

    \begin{center}
    \adjustimage{max size = 0.8\linewidth}{/CrDispersionFits_T295K.png}
    \end{center}
	\caption{Fitting of the Cr L$_3$-edge RIXS data measured at $E{\mathrm{in}} = 578.61$~eV with an incident X-ray angle of 15$^\circ$, using LV polarization at $T = 295$~K. (a) Schematic of the scattering geometry. (b–d) RIXS spectra and corresponding fits measured at scattering angles TTH = 90$^\circ$, 120$^\circ$, and 154$^\circ$, respectively. The fits include an anti-Stokes magnon contribution.}
	\label{FigSI:CrDispersionFits_T295K}
\end{figure}

The FWHM values of the single-, double-, and triple-magnon peaks were determined from the RIXS spectrum measured at a scattering angle of TTH = 154$^\circ$, where the single-magnon excitation reaches its highest energy, allowing for a more reliable determination of the peak widths. No significant broadening of the magnon peaks was observed compared with the corresponding fits to the $T = 15$~K RIXS data. The fits are shown in Fig.~\ref{FigSI:CrDispersionFits_T295K}. For TTH = 90$^\circ$, an additional magnon contribution was included in the fit (shown in yellow in Fig.~\ref{FigSI:CrDispersionFits_T295K}b), as discussed in the main text.

\begin{figure}[h!]
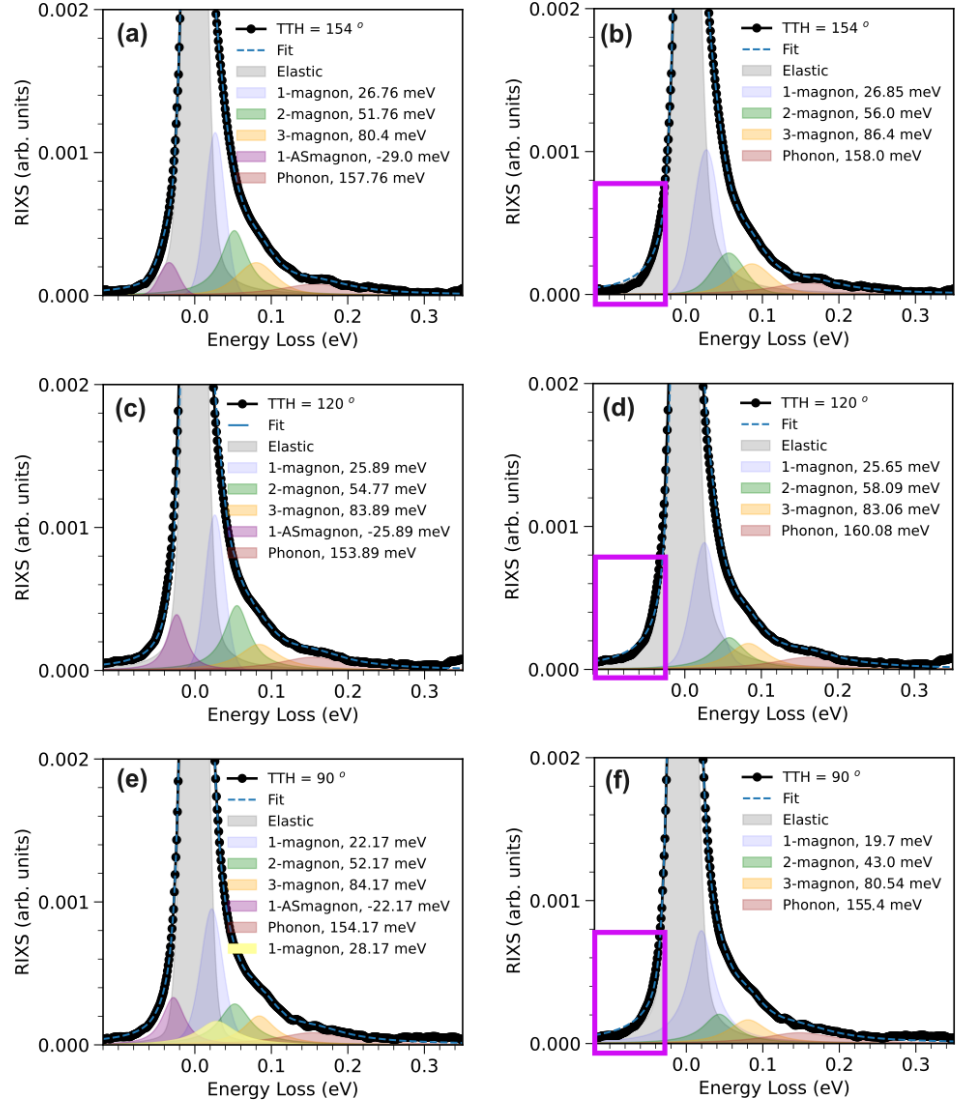

    \begin{center}
    \adjustimage{max size = 0.85\linewidth}{/CrDispersionFits_T295K_FitModels.png}
    \end{center}
	\caption{Fitting of the Cr L$_3$-edge RIXS data measured at $E{\mathrm{in}} = 578.61$~eV with an incident X-ray angle of 15$^\circ$, using LV polarization at $T = 295$~K, employing two fitting models. The left panels show fits including the anti-Stokes magnon contribution, while the right panels show fits without this contribution. The fits are shown for scattering angles TTH = 90$^\circ$, 120$^\circ$, and 154$^\circ$, respectively.}
	\label{Fig:CrDispersionFits_T295K_compareFits}
\end{figure}

In Fig.~\ref{Fig:CrDispersionFits_T295K_compareFits}, we compare two fitting models for the high-temperature RIXS data. The left panels show the fits including the single anti-Stokes magnon contribution, as in Fig.~\ref{FigSI:CrDispersionFits_T295K}, while the right panels show the corresponding fits without this contribution. In the absence of the anti-Stokes magnon, the model fails to reproduce the energy-gain side of the elastic peak for all measured TTH angles, as highlighted by the purple rectangles. This comparison demonstrates that the inclusion of the anti-Stokes magnon contribution is necessary to accurately describe the high-temperature RIXS spectra.

\begin{figure}[h!]
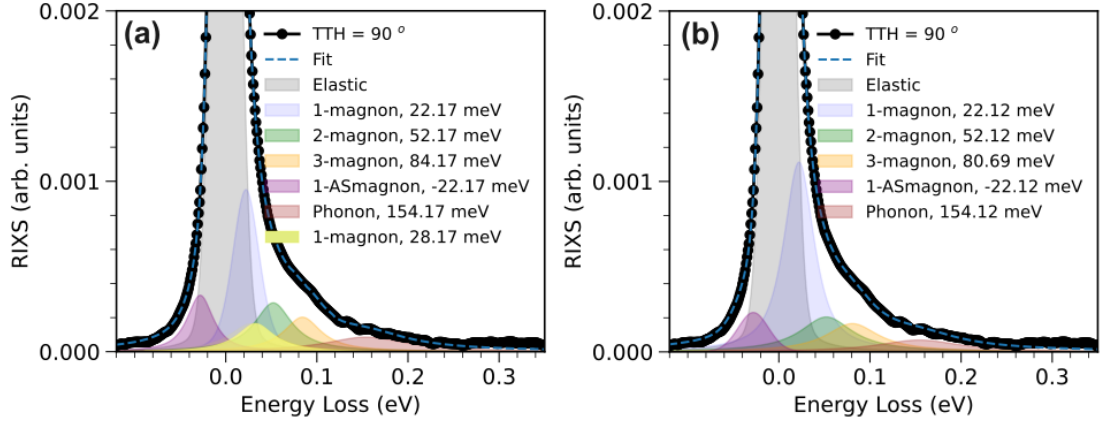

    \begin{center}
    \adjustimage{max size = 0.85\linewidth}{/CrDispersionFits_T295K_6and7Peaks.png}
    \end{center}
	\caption{Fitting of the Cr L$_3$-edge RIXS data measured at $E{\mathrm{in}} = 578.61$~eV with an incident X-ray angle of 15$^\circ$, using LV polarization at $T = 295$~K and a scattering angle of TTH = 90$^\circ$, using two fitting models. (a) Fit including the additional contribution at $\approx 28$~meV. The resulting FWHM values of the single-, double-, and triple-magnon excitations are consistent with those obtained at TTH = 120$^\circ$ and 154$^\circ$. (b) Fit without the $\approx 28$~meV contribution, which requires a substantial broadening of the double- and triple-magnon peaks to reproduce the spectrum. This comparison supports the inclusion of the additional $\approx 28$~meV contribution in the fitting model.}
	\label{FigSI:CrDispersionFits_T295K_6Vs7}
\end{figure}

Finally, in Fig.~\ref{FigSI:CrDispersionFits_T295K_6Vs7}, we compare two fitting models for the TTH = 90$^\circ$ data to assess whether the additional contribution at $\approx 28$~meV is required to describe the spectrum. Figure~\ref{FigSI:CrDispersionFits_T295K_6Vs7}a shows the fit including this contribution. In this case, the single-magnon excitation is found at 22.17~meV with a FWHM of 17.19~meV, while the double- and triple-magnon excitations are located at 52.17 and 84.17~meV, with FWHM values of 25.21 and 24.78~meV, respectively. These linewidths are consistent with those obtained from the TTH = 120$^\circ$ and 154$^\circ$ spectra.

In contrast, when the $\approx 28$~meV contribution is omitted (Fig.~\ref{FigSI:CrDispersionFits_T295K_6Vs7}b), substantially broader multi-magnon peaks are required to reproduce the spectrum. The single-magnon excitation is found at 22.12~meV with a FWHM of 19.93~meV, while the double- and triple-magnon excitations occur at 52.12 and 80.69~meV, with FWHM values of 38.32 and 40.53~meV, respectively. The pronounced increase in the linewidths of the double- and triple-magnon contributions, which is inconsistent with the linewidths obtained at the other scattering angles, indicates that the additional $\approx 28$~meV contribution is required to provide a physically consistent description of the TTH = 90$^\circ$ RIXS spectrum.

\clearpage

\section{Extended data}
\subsection{X-ray absorption spectroscopy and angular dependence}

\begin{figure}[h!]
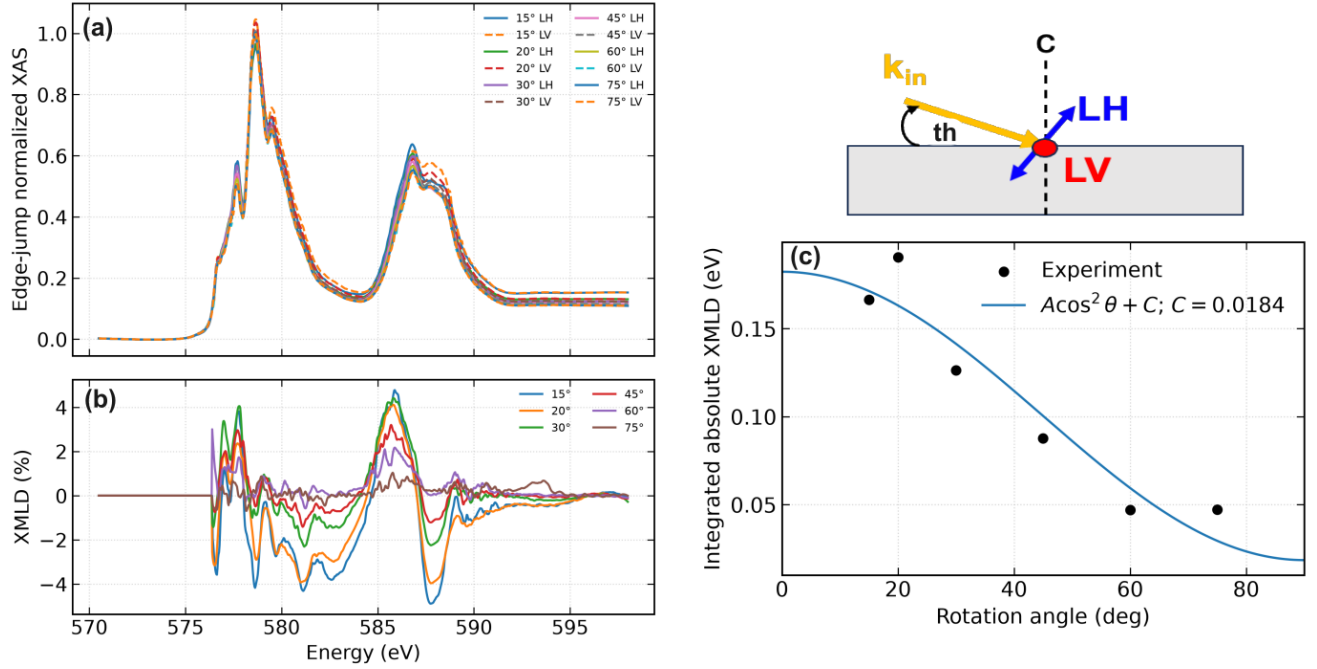

    \begin{center}
    \adjustimage{max size =\linewidth}{/XMLD_ThetaRotation.png}
    \end{center}
	\caption{Angular dependence of the Cr L$_{2,3}$-edge XMLD. (a) XAS spectra measured as a function of the incidence angle $\theta$ using linear horizontal (LH, solid lines) and linear vertical (LV, dashed lines) polarization. (b) Corresponding XMLD spectra, showing an increase in the XMLD amplitude with decreasing $\theta$. (c) Absolute value of the integrated XMLD intensity as a function of $\theta$. The solid line shows a fit to a $\cos^2\theta$ dependence. The maximum XMLD intensity at $\theta=0^\circ$ is consistent with a preferentially out-of-plane orientation of the Néel vector.}
	\label{FigSI:CrXMLD_ThetaRotation}
\end{figure}
We measured the angular dependence of the XMLD signal to determine the orientation of the Néel vector. Figure~\ref{FigSI:CrXMLD_ThetaRotation}a shows the XAS spectra measured as a function of the incidence angle $\theta$ using linear horizontal (LH, solid lines) and linear vertical (LV, dashed lines) polarization. The corresponding XMLD spectra are shown in Fig.~\ref{FigSI:CrXMLD_ThetaRotation}b, revealing a clear increase in the XMLD amplitude with decreasing $\theta$. Figure~\ref{Fig:XMLDTempDependence}b shows the absolute value of the integrated XMLD intensity as a function of $\theta$, together with a fit to its angular dependence. The XMLD intensity follows a $\cos^2\theta$ dependence, as expected for a cubic crystal field. The maximum XMLD intensity at $\theta=0^\circ$ indicates that the Néel vector is preferentially oriented along the out-of-plane direction.

\subsection{Temperature dependence of X-ray magnetic linear dichroism}

\begin{figure}[h!]
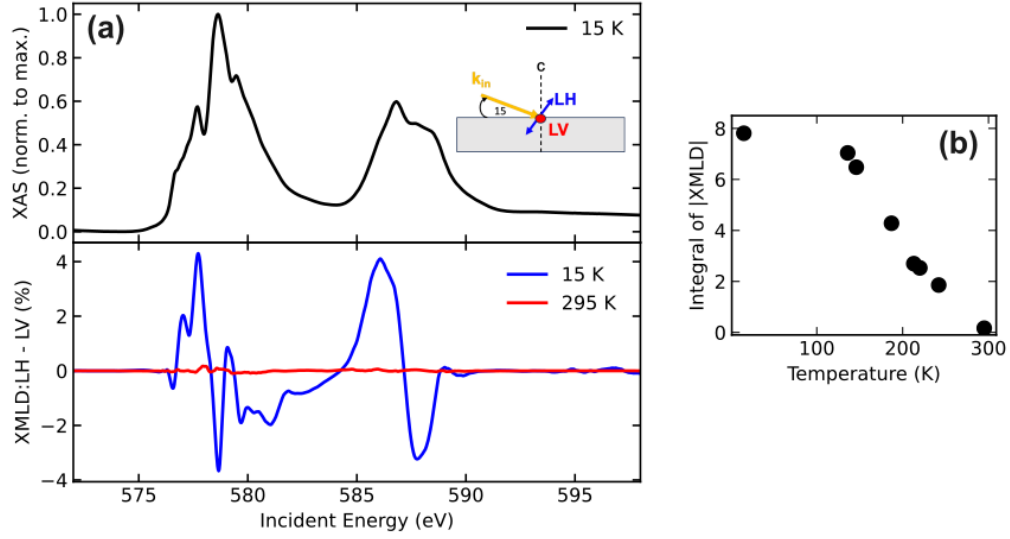

    \begin{center}
    \adjustimage{max size = 0.75\linewidth}{/XMLDResults.png}
    \end{center}
	\caption{(a) Cr L$_{2,3}$-edge XAS and XMLD signals at the minimum (15~K, blue) and maximum (295~K, red) temperatures measured. (b) The absolute value of the integrated area under the Cr L$_{2,3}$-edge XMLD signal as a function of temperature.}
	\label{Fig:XMLDTempDependence}
\end{figure}

The XAS spectra and corresponding XMLD signals measured at the lowest (15~K) and highest (295~K) temperatures are shown in Fig.~\ref{Fig:XMLDTempDependence}a. To determine the Néel temperature, we measured the XMLD as a function of temperature. Figure~\ref{Fig:XMLDTempDependence}b shows the absolute value of the integrated XMLD intensity as a function of temperature. The XMLD intensity progressively decreases with increasing temperature and vanishes at approximately 295~K, indicating the loss of long-range antiferromagnetic order and placing the Néel temperature at approximately $T_\mathrm{N} \approx 295$~K. This observation is consistent with previous reports showing that strain and finite-size effects can substantially modify the magnetic ordering temperature of rare-earth chromite perovskites, $R$CrO$_3$ ($R$ = rare earth), with reported ordering temperatures spanning approximately 120--300~K.

\subsection{Comparison between high and low temperature RIXS measurement}

\begin{figure}[h!]
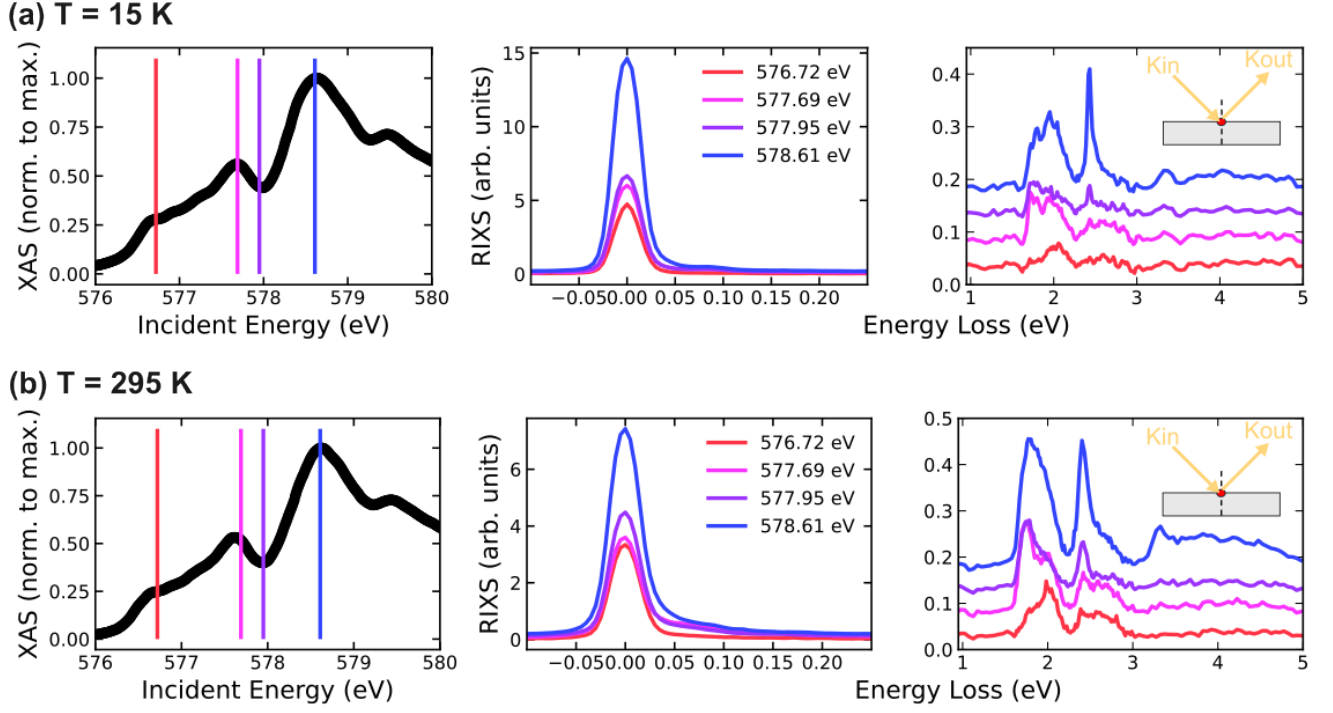

    \begin{center}
    \adjustimage{max size = 1\linewidth}{/RIXSMaps_CompareTemp.png}
    \end{center}
	\caption{Temperature dependence of the Cr L$_3$-edge RIXS energy maps. The scattering geometry is illustrated in the inset. The left panels show the XAS spectra measured at the corresponding temperatures, with vertical lines indicating the incident energies at which the RIXS spectra were acquired. The middle panels show an enlarged view of the low-energy-loss region, while the right panels display the extended energy-loss range containing the $dd$ excitations. (a) RIXS energy map measured at $T = 15$~K. (b) RIXS energy map measured at $T = 295$~K.}
	\label{FigSI:RIXSEnergyMaps}
\end{figure}

Figure~\ref{FigSI:RIXSEnergyMaps} shows the RIXS energy maps measured at $T = 15$~K (panel a) and $T = 295$~K (panel b). At high temperature, the $dd$ excitations become noticeably broader and the individual spectral features are less well resolved than at low temperature. In particular, the sharp feature observed at an energy loss of approximately 2.42~eV at $T = 15$~K exhibits a pronounced broadening at $T = 295$~K.

\begin{figure}[h!]
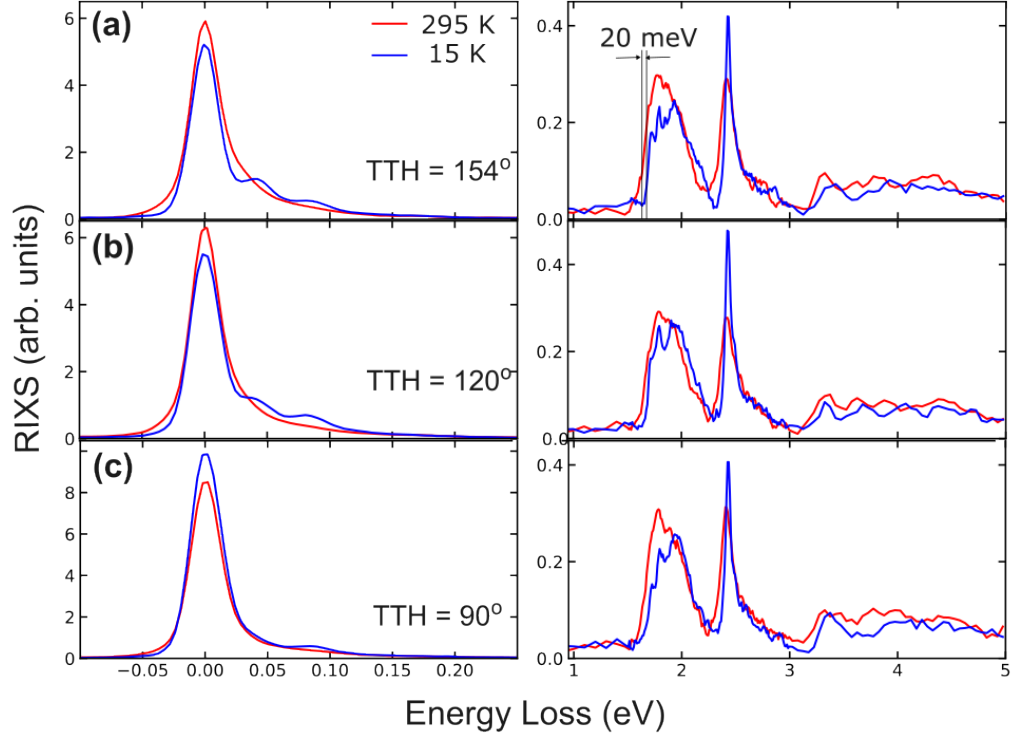

    \begin{center}
    \adjustimage{max size = 0.75\linewidth}{/RIXS_Tempdepend_Magnon-dd.png}
    \end{center}
	\caption{Temperature dependence of the Cr L$_3$-edge RIXS spectra measured at (a) TTH = 154$^\circ$, (b) 120$^\circ$, and (c) 90$^\circ$. The spectra measured at $T = 15$~K (blue) and $T = 295$~K (red) are overlaid for direct comparison. The left panels show an enlarged view of the low-energy region. The right panel shows the extended energy-loss range shows the temperature evolution of the $dd$ excitations.}
	\label{FigSI:RIXSDispersion}
\end{figure}

Figure~\ref{FigSI:RIXSDispersion} compares the RIXS spectra measured at $T = 15$~K (blue) and $T = 295$~K (red) for TTH = 154$^\circ$, 120$^\circ$, and 90$^\circ$. The left panels show an enlarged view of the low-energy region. At high temperature, the magnon excitations are damped and a pronounced increase in spectral weight appears on the energy-gain side of the elastic peak. Importantly, this additional spectral weight exhibits a clear momentum dependence. At TTH = 154$^\circ$, a distinct side peak is visible on the energy-gain side, which progressively shifts towards the elastic line upon decreasing the scattering angle to TTH = 90$^\circ$. This evolution mirrors the dispersion of the corresponding Stokes single-magnon excitation and provides direct evidence for the presence of an anti-Stokes magnon contribution in the high-temperature RIXS spectra.

The temperature dependence of the $dd$ excitations is also apparent across all measured scattering angles. In particular, the $\sim 20$~meV energy shift of the $dd$ excitations is consistently observed for all TTH values, together with the pronounced broadening of the sharp feature at approximately 2.42~eV.

\clearpage
\section{Superconducting quantum interference device (SQUID) magnetometry}

\begin{figure}[h!]
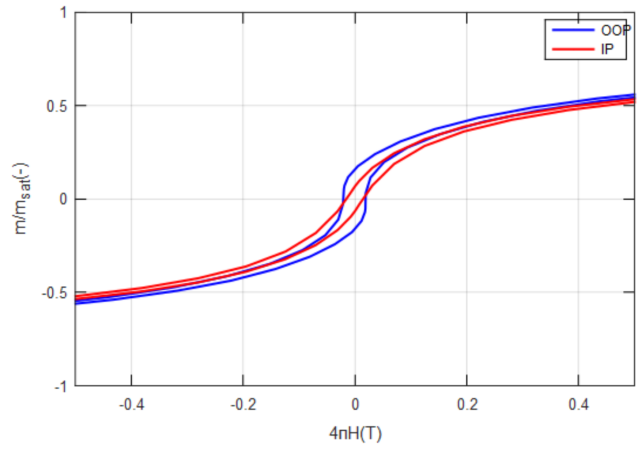

    \begin{center}
    \adjustimage{max size = 0.5\linewidth}{/SQUID.png}
    \end{center}
	\caption{Magnetic-field dependence of the LaCrO$_3$/Nb:SrTiO$_3$ thin film at T = 15~k. SQUID magnetometry measured with the magnetic field applied in-plane (IP, red) and out-of-plane (OOP, blue), showing a weak magnetic anisotropy consistent with the out-of-plane Néel-vector orientation inferred from XMLD.}
	\label{FigSI:SQUID}
\end{figure}

Figure~\ref{FigSI:SQUID} shows the magnetic-field dependence of the magnetization measured at 15~K with the magnetic field applied in-plane (IP) and out-of-plane (OOP), using the same orientations as in the XMLD measurements. Both configurations exhibit a predominantly weak, non-saturating magnetic response, consistent with the antiferromagnetic character of LaCrO$_3$. A small hysteretic contribution is observed around zero field, with a somewhat more pronounced response in the OOP configuration. The IP and OOP curves also exhibit a weak difference in their high-field slopes, indicating a finite magnetic anisotropy. While the absolute magnitude of the magnetic response contains contributions from the Nb:SrTiO$_3$ substrate and therefore cannot be uniquely assigned to the 10-nm LaCrO$_3$ film in the absence of a bare-substrate reference, the observed anisotropy is consistent with the preferential out-of-plane orientation of the Néel vector inferred from the angular dependence of the XMLD. In particular, the XMLD measurements show a $\cos^2\theta$ dependence with maximum intensity for $\theta=0^\circ$, identifying the out-of-plane direction as the preferred orientation of the Néel vector. The weak anisotropy observed in the SQUID measurements is consistent with this magnetic configuration, providing complementary evidence for an out-of-plane antiferromagnetic order parameter.

\clearpage
\section{Calculations}
\subsection{Crystal-field multiplet calculations}

Crystal-field multiplet calculations were performed to describe the Cr $L_{2,3}$-edge XAS and RIXS spectra. The calculations consider a single Cr$^{3+}$ ion with a $3d^3$ electronic configuration in the crystal field generated by the surrounding oxygen octahedron. The electronic structure was obtained from a configuration-interaction calculation including the intra-atomic $3d$--$3d$ Coulomb interaction, the crystal-field interaction, and the spin--orbit coupling. For the RIXS process, the $2p$--$3d$ Coulomb and exchange interactions and the $2p$ spin--orbit coupling were additionally included in the intermediate state.

The local symmetry of the Cr site was described using a tetragonal $D_{4h}$ crystal field to account for the distortion of the CrO$_6$ octahedron. The crystal-field Hamiltonian was parameterized in terms of 10Dq , Ds, and Dt, where 10Dq describes the dominant octahedral splitting and Ds and Dt account for the tetragonal distortion. The crystal-field parameters were adjusted to reproduce the experimentally observed $dd$ excitation energies according the following procedure. The first quartet excitation gives the effective 10Dq. In LaCrO$_3$ this is 1.99 eV. Then, a Tanabe Sugano diagram is calculated with 10D$_q$ = 1.99 eV Dt = Ds = 0.0 eV and slater integrals values 80\% of the Hartree Fock values as a sanity check. After empirically determining 10Dq values, Dt and Ds values are extracted by optimising the fitting of the XMLD calculation with experimental XMLD spectrum. The energy of quartet states  are expressed in terms of Racah B and doublet states in Racah B and C. The energy difference between the first and second quartet state is B dependent hence this parameter is optimized. The final values we found for the T = 15 K data are reported in Table \ref{TabSI:MultipletParameters}.

The Cr $L_{2,3}$-edge XAS and RIXS spectra were calculated for the dipole-allowed transitions
$2p^63d^3 \rightarrow 2p^53d^4$. The RIXS intensity was calculated using the Kramers--Heisenberg formalism, taking into account the experimental incident polarization and scattering geometry using the code Quanty. Atomic Slater integrals were obtained from Hartree--Fock calculations and reduced to account for screening and covalency effects. The calculated spectra were subsequently broadened to account for the finite core-hole lifetime and experimental energy resolution.

\begin{table}[h]
\begin{tabular}{lc}
\hline
Parameter & Value (eV) \\
\hline
\hline
$10Dq$ & 1.99 \\
$D_s$ & 0.05 \\
$D_t$ & 0.005 \\
$F^2_{dd}$ & 4.96 \\
$F^4_{dd}$ & 6.08 \\
$F^2_{pd}$ & 6.53  \\
$G^1_{pd}$ & 3.26 \\
$G^3_{pd}$ & 2.38 \\
$\zeta_{3d}$ & 0.035 \\
$\zeta_{2p}$ & 5.667 \\
$J_{exch}$ & 0.035 \\
\hline
\end{tabular}
\caption{Parameters used in the Cr$^{3+}$ crystal-field multiplet calculations.}
\label{TabSI:MultipletParameters}
\end{table}

We re-optimized the calculations for the T = 295 K data. We found that the main modification required to explain these data is to increase the Slater integrals to $F^2_{dd}$ = \SI{5.07}{eV} and $F^4_{dd}$ = \SI{6.28}{eV}.

\subsection{Linear spin-wave and multimagnon calculations}

The magnetic excitation spectrum of LaCrO$_3$ was calculated within linear spin-wave theory using SpinW \cite{Toth2015SpinW}. The Cr$^{3+}$ ions were described as localized spins with $S=3/2$ using an isotropic nearest-neighbour Heisenberg Hamiltonian,
\begin{equation}
\mathcal{H}=J_1\sum_{\langle i,j\rangle}\mathbf S_i\cdot\mathbf S_j ,
\end{equation}
with antiferromagnetic exchange $J_1=4.8$~meV. A primitive two-sublattice magnetic cell was used, with antiparallel Cr moments corresponding to collinear G-type antiferromagnetic order. No magnetic anisotropy, Dzyaloshinskii--Moriya interaction, or further-neighbour exchange was included.

For the multimagnon calculations, the positive-energy one-magnon branches were evaluated on a uniform $32\times32\times32$ momentum grid. The two- and three-magnon spectra were constructed within a non-interacting-magnon approximation, in which the energy of a multimagnon state is the sum of the constituent one-magnon energies. The corresponding unweighted joint densities of states were calculated as:

\begin{equation}
A_2(\mathbf Q,\omega)
=
\frac{1}{N_{\mathbf q}}
\sum_{\mathbf q,\nu,\mu}
\delta\!\left[
\omega-\varepsilon_\nu(\mathbf q)
-\varepsilon_\mu(\mathbf Q-\mathbf q)
\right],
\end{equation}
and
\begin{equation}
A_3(\mathbf Q,\omega)
=
\frac{1}{N_{\mathbf q}^2}
\sum_{\mathbf q_1,\mathbf q_2}
\sum_{\nu,\mu,\lambda}
\delta\!\left[
\omega-\varepsilon_\nu(\mathbf q_1)
-\varepsilon_\mu(\mathbf q_2)
-\varepsilon_\lambda(\mathbf Q-\mathbf q_1-\mathbf q_2)
\right].
\end{equation}
Magnon--magnon interactions, bound states, and interaction-induced renormalization were not included. The calculated spectra were finally broadened with a 1~meV Gaussian for visualization.

At finite temperature, the magnon occupation was described by the Bose--Einstein distribution,
\begin{equation}
n(\varepsilon,T)
=
\frac{1}{\exp(\varepsilon/k_{\mathrm B}T)-1}.
\end{equation}
The corresponding creation channels were weighted by factors
$1+n(\varepsilon,T)$, whereas annihilation channels were weighted by
$n(\varepsilon,T)$. The finite-temperature three-magnon creation
channel therefore carries the weight
$(1+n_1)(1+n_2)(1+n_3)$.

In addition, a mixed three-magnon channel involving the creation of two
magnons and the annihilation of one thermally populated magnon was
calculated as
\begin{equation}
A_{++-}(\mathbf Q,\omega;T)
=
\frac{1}{N_{\mathbf q}^2}
\sum_{\mathbf q_1,\mathbf q_2}
\sum_{\nu,\mu,\lambda}
(1+n_1)(1+n_2)n_3
\,
\delta\!\left[
\omega
-\varepsilon_\nu(\mathbf q_1)
-\varepsilon_\mu(\mathbf q_2)
+\varepsilon_\lambda(\mathbf q_3)
\right],
\end{equation}
where
\begin{equation}
\mathbf q_3=\mathbf q_1+\mathbf q_2-\mathbf Q
\end{equation}
follows from momentum conservation, modulo a reciprocal-lattice
vector, and
$n_i=n[\varepsilon_i,T]$. All momenta were folded periodically onto the
magnetic Brillouin-zone grid. This contribution vanishes in the
$T\rightarrow0$ limit because the annihilation process requires a
thermally occupied initial magnon state.

The thermally activated single-magnon anti-Stokes contribution was
treated analogously, with an intensity proportional to
$n(\varepsilon,T)$, while the Stokes contribution is proportional to
$1+n(\varepsilon,T)$.

For visualization, the energy-binned spectra were convolved with a
Gaussian of 1~meV full width at half maximum. This broadening was
applied only after construction of the JDOS and does not represent an
intrinsic magnon lifetime.

%
\bibliographystyle{unsrt}
\bibliography{Reference}

\end{document}